\documentclass[conference,10pt]{IEEEtran}

\usepackage{subfigure}
\usepackage{epsf}
\usepackage{amsmath,amssymb}
\usepackage{graphicx}
\usepackage{color}
\usepackage{cite}
\usepackage{multirow,tabularx}
\usepackage{ifthen}
\usepackage{times}
\usepackage{stackrel}
\usepackage{amsfonts}
\usepackage{stfloats}
\usepackage[all]{xy}
\usepackage{url}
\usepackage{framed}

\input{epsf.sty}
\newcommand{\hide}[1]{\ifthenelse{\boolean{false}}{#1}{}}

\newtheorem{theorem}{{\bf Theorem}}
\newtheorem{lemma}{{\bf Lemma}}

\newtheorem{definition}{\bf Definition}

\newcommand{\qed}{\nobreak \ifvmode \relax \else
      \ifdim\lastskip<1.5em \hskip-\lastskip
      \hskip1.5em plus0em minus0.5em \fi \nobreak
      \vrule height0.75em width0.5em depth0.25em\fi}

\newcommand{\iid}{{i.i.d.}}

\newcommand{\expect}[1]{\mathbb{E}\left[#1\right]} % expectation operator
\newcommand{\prob}[1]{\mathbb{P}\left[#1\right]}

\newcommand{\bsp}{\begin{slide*}}
\newcommand{\esp}{\end{slide*}}
\newcommand{\bsl}{\begin{slide}}
\newcommand{\esl}{\end{slide}}

\begin{document}

\IEEEoverridecommandlockouts

\title{Capacity and Delay Scaling for Broadcast Transmission in Highly Mobile Wireless Networks}
\author{Rajat Talak, Sertac Karaman, and Eytan Modiano
\thanks{The authors are with the Laboratory for Information and Decision Systems (LIDS) at the Massachusetts Institute of Technology (MIT), Cambridge, MA. {\tt \{talak, sertac, modiano\}@mit.edu}}
%\thanks{This work first appeared in MobiHoc 2017~\cite{talak17_Mobihoc}.}
}

\maketitle
\setcounter{page}{1}

\begin{abstract}
We study broadcast capacity and minimum delay scaling laws for highly mobile wireless networks, in which each node has to disseminate or broadcast packets to all other nodes in the network. In particular, we consider a cell partitioned network under the simplified independent and identically distributed (IID) mobility model, in which each node chooses a new cell at random every time slot. We derive scaling laws for broadcast capacity and minimum delay as a function of the cell size. We propose a simple first-come-first-serve (FCFS) flooding scheme that nearly achieves both capacity and minimum delay scaling. Our results show that high mobility does not improve broadcast capacity, and that both capacity and delay improve with increasing cell sizes. In contrast to what has been speculated in the literature we show that there is (nearly) no tradeoff between capacity and delay. Our analysis makes use of the theory of Markov Evolving Graphs (MEGs) and develops two new bounds on flooding time in MEGs by relaxing the previously required expander property assumption.
\end{abstract}

\section{Introduction}
\label{sec:intro}
We study all-to-all broadcast capacity and delay scaling behavior in mobile wireless networks. Interest in mobile wireless networks has increased in recent years due to the emergence of autonomous aerial vehicle (UAV) networks. Dense networks of small UAVs are being used in a wide range of applications including product delivery, disaster and environmental monitoring, surveillance, and more~\cite{VJ2012_MAV, VJ2014_MAV, Kushleyev2013_MAVQuadrotors, FANETs2014, gupta_uavnet}. Our work is motivated by the need to disseminate timely control information in such networks~\cite{FANETs2014, gupta_uavnet, hayat_uavnet, talak_cdc16}.
An important communication operation that needs to be performed in exchanging safety critical information is that of all-to-all broadcast, where each vehicle or node broadcasts its current state or location information to all other vehicles in its vicinity.

\begin{figure}
\centering
\includegraphics[width=0.75\linewidth]{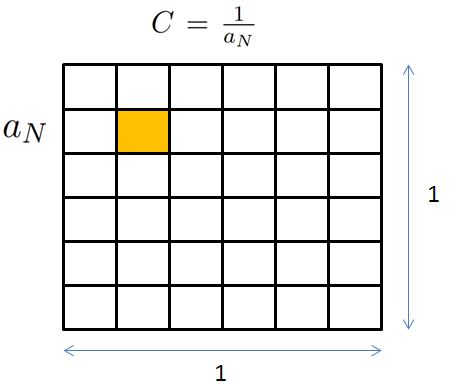}
\caption{Network partitioned into $C = \frac{1}{a_N}$ cells. Each cell of area $a_N$.}
\label{fig:sys_model}
\end{figure}

We consider a cell partitioned network with $N$ nodes, shown in Figure~\ref{fig:sys_model}, in which a unit square is partitioned into $C$ cells. Due to interference, only a single packet transmission can take place in the cell at a given time, and all other nodes in the cell can correctly receive the packet. Different cells can have simultaneous packet transmissions. This simple model captures the essential features of interference and helps obtain key insights into its impact on throughput and delay~\cite{modiano, dev_shah06, goldsmith}. We consider IID mobility, where, at the end of every slot, each node chooses a new cell uniformly at random.
This mobility model was used in~\cite{modiano, motioncast2011} to capture the impact of high mobility, and the resultant intermittent network connectivity, on throughput and delay. Moreover, this model serves as a good model for UAV networks where rapid mobility and intermittent connectivity are common~\cite{gupta_uavnet, hayat_uavnet, FANETs2014}.

% \subsection{Contributions}
%

We study all-to-all broadcast capacity and delay scaling as a function of node density. Here, capacity is defined as the maximum rate at which each node can transmit packets to all other nodes in the system and delay as the average time taken by a packet to reach every node in the system.
We say that a network is dense if the number of vehicles or nodes per cell is increasing with $N$, and sparse otherwise. Thus, if the cell size grows as $cN^{-\alpha}$, for some $c > 0$, then the network is {\em dense} for $0 < \alpha < 1$ and {\em sparse} for $\alpha \geq 1$.

We show that as the network gets more dense the all-to-all broadcast capacity increases to reach a maximum scaling of $1/N$. Interestingly, delay decreases as the network gets denser. In fact, both, capacity and delay attain their best scaling in $N$ when the cell size is just smaller than order $1/N$, i.e., when $\alpha = 1 - \epsilon$ for a small positive $\epsilon$.
%
% Since wireless nodes are often power constrained, this result implies that it suffices for each vehicle to expend transmission power proportional to $1/N^{1-\epsilon}$ for a small $\epsilon > 0$, as any higher transmission power does not yield any further gain. Thus, the total transmission power used by the system is $N^{\epsilon}$, and $\epsilon > 0$ can be made as small as possible.
%
We further note that the best per-node capacity scaling of $1/N$ is the same as that can be achieved in a static wireless network, thus, mobility does not improve network capacity. This is in contrast to the unicast case where it was shown in~\cite{tse02} that mobility improves capacity. Our scaling results are summarized in Table~\ref{tbl:rate}.
\begin{table}
\caption{Capacity and Average Delay}
\label{tbl:rate}
\begin{center}
\begin{tabular}{ ccc }
\hline
 & \multicolumn{2}{c}{\textbf{Capacity}} \\
\hline
 & Upper bound & FCFS flooding\\
 & (Theorem~\ref{thm:capacity}) & (Eqn.~\eqref{eq:fcfs_flooding_cap}) \\
\hline
Sparse: $\alpha \geq 1$ & $\frac{1}{N^\alpha}$ & $\frac{1}{N^\alpha}\frac{1}{\log N}$ \\
%\hline
Dense: $0< \alpha < 1$ & $\frac{1}{N}$ & $\frac{1}{N}\frac{1}{\log \log N}$\\
%\hline
%\hline
\hline
 & \multicolumn{2}{c}{\textbf{Average Delay}} \\
\hline
 & Lower bound & FCFS flooding\\
 & (Theorem~\ref{thm:delay_lb}) & (Eqns.~\eqref{eq:fcfs_flooding_delay} and~\eqref{eq:un}) \\
\hline
Sparse: $\alpha \geq 1$ & $N^{\alpha -1} \log N$ & $N^{\alpha -1} \log N$ \\
%\hline
Dense: $0< \alpha < 1$ & $1$ & $\log \log N$\\
\hline
\end{tabular}
\end{center}
\end{table}

We propose a simple first-come-first-serve (FCFS) flooding scheme that achieves capacity scaling, up to a $\log N$ factor from the optimal when the network is sparse and up to a $\log \log N$ factor from the optimal when the network is dense. The FCFS flooding scheme also achieves the minimum delay scaling when the network is sparse, and up to a factor of $\log \log N$ from minimum delay when the network is dense. Thus, nearly optimal throughput and delay scaling is achieved simultaneously. % This also implies that when each vehicle expends transmission power proportional to $1/N^{1-\epsilon}$, for a small $\epsilon > 0$, the FCFS flooding scheme achieves capacity and minimum delay scaling up to a factor of $\log \log N$. This nearly optimal throughput and delay is achieved simultaneously.

The IID mobility model was analyzed for unicast and multicast operations in~\cite{modiano} and~\cite{motioncast2011}, respectively, using standard probabilistic arguments. In contrast, we use the abstraction of Markov evolving graphs (MEG), and flooding time bounds for MEGs~\cite{clementi11_MEG}. An MEG is a discrete time Markov chain with state space being a collection of graphs with $N$ nodes. An MEG of the IID mobility model can be constructed by drawing an edge between two nodes in the same cell and viewing the network as a graph at each time step. Flooding time, is then, the time it takes for a single packet to reach all nodes from a single source node.

A flooding time bound for MEGs was derived in~\cite{clementi11_MEG}. It relied on an expander property which states that whenever $m$ nodes have the packet then in the next slot at least $km$ new nodes will receive the packet with high probability, for some $k > 0$. However, this strong requirement does not always hold. For example, when the IID mobility model is sparse, this expander property cannot be guaranteed. We derive two new bounds on flooding time in MEGs by relaxing the strong expander property requirements imposed in~\cite{clementi11_MEG}. These new bounds are of independent theoretical interest. This work first appeared in MobiHoc 2017~\cite{talak17_Mobihoc}.

\subsection{Previous Work}
In~\cite{talak_cdc16}, we considered the impact of wireless interference constraints on the ability to exchange timely control information in UAV networks. We showed that, in guaranteeing location awareness of other vehicles in the networks, wireless interference constraints can limit mobility of aerial vehicles in such networks. This result motivates us to study the delay and capacity scalings of all-to-all broadcast in mobile wireless networks.

Broadcast has been studied before in the contexts of disseminating data packets in wireless ad-hoc networks~\cite{li08_bc, taliv2006}, sensor information in sensor networks, and in exchanging intermediate variables in distributed computing~\cite{tsi_parallel}.  Scaling laws for capacity and delay in wireless networks have received significant attention in the literature. Capacity scaling for unicast traffic, in which each node sends packets to only one other destination node, was analyzed in~\cite{gupta_kumar00, kulkarni_pv}. It was shown that the capacity scales as $1/\sqrt{N \log N}$ with increasing $N$. Minimum delay scaling for the static unicast network was analyzed in~\cite{dev_shah06}, where it was also shown that it is not possible to simultaneously achieve minimum delay and capacity. This implied a tradeoff between capacity and delay.
In~\cite{tse02}, it was shown that if the nodes were mobile, then a constant per node capacity that does not diminish with $N$ can be achieved. The seminal works of~\cite{gupta_kumar00} and~\cite{tse02} led to the analysis of capacity and delay scaling under various mobility models including IID~\cite{modiano}, Markov~\cite{dev_shah06}, Brownian motion~\cite{ravi_ness06}, and Random Waypoint~\cite{sharma_ravi05}. Capacity-delay tradeoffs were observed in each of these settings.

Broadcast has been studied in static wireless networks in~\cite{shakkottai_srikant2010_multicast_cap, taliv2006, li08_bc, 2006bc_scaling}. It was shown that the per-node broadcast capacity scales as $1/N$ in static wireless networks~\cite{taliv2006}. However, to the best of our knowledge, optimal delay scalings for static broadcast has not been analyzed. In~\cite{motioncast2011}, the authors conjectured a capacity-delay tradeoff for multicast, and by implication for broadcast as a special case, under IID mobility. However, in this paper, we show that there is nearly no capacity-delay tradeoff for broadcast. In particular, we propose a scheme that (nearly) achieves both capacity and minimum delay, which is up to a $\log \log N$ factor when the network is dense and up a $\log N$ factor when the network is sparse. Moreover, we show that the capacity scaling does not improve with mobility, unlike in the unicast case~\cite{tse02}.

Although, throughput and delay scalings have been investigated under various communication operations and mobility models for the past 15 years, the same problem under broadcast has not been thoroughly analyzed even for the simplest IID mobility model. In~\cite{motioncast2011}, delay bounds were obtained for multicast, however, these bounds are very weak when applied to the all-to-all broadcast operation. By using and extending the theory of MEGs developed in~\cite{clementi11_MEG} we are able to obtain tight bounds on delay.

Flooding time bounds on MEG have been used for various network models in~\cite{clementi11_MEG, Clementi2013_manets, clementi2015}. To the best of our knowledge, this is the first time that these techniques are being used in the mobility setting. Moreover, the new bounds derived in Section~\ref{sec:meg} could be of independent interests and can also be applied to models considered in~\cite{clementi11_MEG, Clementi2013_manets, clementi2015}.

\subsection{Organization}
The paper is organized as follows. In Section~\ref{sec:model} we describe the system model, and in Section~\ref{sec:cap_delay} we derive bounds on capacity and minimum delay. In Section~\ref{sec:meg}, we summarize the flooding time upper bound result of~\cite{clementi11_MEG}, and derive two new upper bounds on flooding time for MEGs. In Section~\ref{sec:flood_time}, we apply these results to our setting and, in Section~\ref{sec:fcfs_flood}, we use it to analyse the FCFS flooding scheme. We propose a single-hop scheme in Section~\ref{sec:single_hop} that achieves capacity for a sparse network. We conclude in Section~\ref{sec:conclusion}.

\section{System Model}
\label{sec:model}
Consider the network of Figure~\ref{fig:sys_model} with $N$ nodes that are uniformly distributed over a unit square. The size of each cell is $a_N = \frac{1}{C} = c N^{-\alpha}$, for some $\alpha > 0$ and $c > 0$.%\footnote{We restrict $a_N$ to be of the form $c N^{-\alpha}$ only for clarity of presentation. The results, and their proofs, can be easily generalized to any other $a_N$.}
We consider a slotted time system, with the duration of each slot normalized to unity. The duration of each slot is sufficient to complete the transmission of a single packet.
We use the IID mobility model of~\cite{modiano} in which each node, at the end of every slot, chooses a new cell/location uniformly at random, and independent of other node's locations.

Packets arrive at each node according to a Poisson process, at rate $\lambda$. Note that the arrivals happen over continuous time, and therefore, two or more packets can arrive during a slot.

In this paper we make extensive use of order notation. For infinite sequences $\left\{ a_N\right\}$ and $\left\{ b_N\right\}$, $a_N = O\left(b_N\right)$ implies $\lim_{N \rightarrow \infty} \frac{a_N}{b_N} \leq c_1$ for some $c_1 > 0$ and $a_N = \Theta\left(b_N\right)$ implies $a_N = O\left(b_N\right)$ and $b_N = O\left(a_N\right)$. We write $a_N \leq_N b_N$ if there exists a $N_0 \geq 1$ such that for all $N \geq N_0$ we have $a_N \leq b_N$. Positive constants are denoted by $c_1, c_2 \ldots$. %, and binomial, geometric, Bernoulli random variables are denoted as %$\prob{\cdot}$ and $\expect{\cdot}$ denote probability and the expected value, respectively. $X \sim \text{Bin}(n,p)$ implies that $X$ is a random variable with binomial distribution with parameters $n$ and $p$, i.e., $\prob{X = k} = {n \choose p} p^k (1-p)^{n-k}$ for all $k \in [n]$. Similarly, $X \sim \text{Geo}(p)$ implies that $X$ is a geometrically distributed random variable with parameter $p$, i.e., $\prob{X=k} = p \left( 1 - p\right)^{k-1}$ for all $k \geq 1$, and $X \sim \text{Ber}(p)$ implies that $X$ is a Bernoulli random variable such that $\prob{X = 1} = p = 1 - \prob{X = 0}$. Cardinality of a set $I$ is denoted by $|I|$.

\section{Fundamental Limits: Capacity and Minimum Delay}
\label{sec:cap_delay}
We now obtain upper-bound on rate $\lambda$ and a lower-bound on achievable delay.

\subsection{Capacity}
\label{sec:cap}
Each node receives an inflow of packets at rate $\lambda$, and each of these packets have to be broadcast to all other nodes in the network. A communication scheme is said to achieve a rate of $\lambda$ if at this arrival rate the average number of backlogged packets in the network does not increase to infinity. The capacity of the network is the maximum achievable rate. We start with a simple upper-bound on the capacity.
\begin{framed}
\begin{theorem}
\label{thm:capacity}
The achievable rate $\lambda$ is bounded by
\begin{align}
\lambda &\leq \frac{1}{2(N-1)} \left( 1 - \left(1 - a_N\right)^{N-1}\right) \\
        &=  \left\{ \begin{array}{ll}
            \Theta\left( \frac{1}{N^\alpha}\right) & \text{if}~\alpha \geq 1~\text{(sparse)} \\
            % \Theta\left( \frac{1}{N}\right) & \text{if}~\alpha = 1~\text{(critical)}\\
            \Theta\left( \frac{1}{N}\right) & \text{if}~0 < \alpha < 1~\text{(dense)}
            \end{array} \right..
\end{align}
\end{theorem}
\end{framed}

\begin{IEEEproof}
For an intuitive argument, consider a scheme that achieves a rate of $\lambda$. Then the average number of packet receptions per slot must be at least $N(N-1)\lambda$ under this scheme, because there are $(N-1)$ destinations for each of the $N$ sources. However, the total number of receptions per slot cannot be more than the average number of nodes in each cell, across all cells. Thus,
\begin{align}
N(N-1) \lambda &\leq \text{average no.\ receptions in each slot} \\
&\approx C \sum_{k=2}^{N} k \prob{k~\text{nodes in a cell}} \label{mmm}\\
&= \frac{1}{a_N} \sum_{k=2}^{N} k {N \choose k} a_{N}^{k} \left( 1 - a_N \right)^{N-k} \\
&= N \left\{ 1 - \left( 1 - a_N \right)^{N-1}\right\}.
\end{align}
In~\eqref{mmm}, the summation starts from $k=2$ as there must be at least two nodes in a cell to have a transmission. The above intuition turns out to be true. Scaling law of the upper bound is then obtained by substituting $a_N = c N^{-\alpha}$.
The complete proof is given in Appendix~\ref{pf:thm:capacity}.
\end{IEEEproof}

This capacity upper bound is in fact achievable. The single-hop scheme in Section~\ref{sec:single_hop} achieves capacity when the network is sparse and the FCFS flooding scheme in Section~\ref{sec:fcfs_flood} achieves capacity, up to a $\log \log N$ factor, when the network is dense. Typically, one expects to have larger broadcast capacity with increasing cell sizes, i.e., with decreasing $\alpha$. A larger cell size implies more nodes in a given cell, and hence, more receptions per slot can occur by exploiting the broadcast nature of the wireless medium. Theorem~\ref{thm:capacity}, however, shows that the capacity remains constant at $\Theta\left(\frac{1}{N}\right)$ for $0 < \alpha < 1$. This is because, larger cell sizes also result in fewer transmission opportunities in every slot due to interference. As a result capacity remains constant when $0 < \alpha < 1$.

\subsection{Minimum Delay}
\label{sec:delay}
Another important performance measure is the delay. The delay of a packet is defined as the time from the arrival of the packet to the time the packet reaches all its $N-1$ destination nodes. The delay of a communication scheme is the average delay, averaged over all packets in the network. To obtain a lower-bound on the network's delay performance we define a \emph{single packet flooding} scheme that transmits a single packet to all other nodes in the network. As we show later, this lower-bound provides a fundamental limit on delay.
%A lower bound for delay is the time it takes for a single packet to reach all other nodes with no other packets in the network. The fastest way to do this is via \emph{packet flooding} as described below:

\begin{framed}
\underline{Single packet flooding scheme}: At the beginning of the first slot, only a single node has the packet.
\begin{enumerate}
  \item In every cell, randomly select one packet carrying node to be the transmitter in that slot. If no such node exists in a cell no transmission occurs in that particular cell.

  \item In each cell, the transmitter node (if present) transmits the packet to all other nodes in the cell.

  \item If all nodes have the packet then terminate the process, otherwise repeat from step~1.
\end{enumerate}
\end{framed}
The single packet flooding scheme is clearly the fastest way to disseminate a packet to all nodes in the network. Hence, a lower-bounded is given by the time it takes for a single packet to reach all other nodes under the single packet flooding scheme.

The analysis of the single packet flooding scheme relies on the following observation: if $h$ nodes have the packet at a given time slot then the number of nodes that will receive the packet in the next slot, $N(h)$, is a binomial random variable $\text{Bin}( N-h, 1 - \left( 1 - a_N\right)^h )$.

To see this, let $H = \{1, 2, \ldots h\}$ and $\overline{H} = \{h+1, h+2, \ldots N\}$ denote the set of nodes that have and do not have the packet at a given time slot, respectively. For the node $i$ that has not received the packet, i.e. $i \in \overline{H}$, let $X_i$ be a binary valued random variable that is $1$ if node $i$ receives the packet in the next slot and $0$ otherwise. The probability that the node $i$ does not receive the packet in the next slot is the probability that no node of $H$ lies in the same cell as node $i$. This happens with probability $\left(1-a_N\right)^h$ as locations of node's are independent and identically distributed (\iid). Hence, $\prob{X_i = 0} = \left( 1 - a_N\right)^h$. Also, the $X_i$s are independent across $i \in \overline{H}$ as, again, the node locations are \iid~and uniform. Since $N(h) = \sum_{i \in H} X_i$ the result follows.
We use this to obtain a lower-bound on delay.
\begin{framed}
\begin{theorem}
\label{thm:delay_lb}
Any achievable average delay $\overline{D}$ is lower-bounded by
\begin{equation}
\overline{D} \geq \left\{ \begin{array}{ll}
            \Theta\left( N^{\alpha -1} \log N \right) & \text{if}~\alpha \geq 1~\text{(sparse)} \\
            % \Theta\left( \log N \right) & \text{if}~\alpha = 1~\text{(critical)} \\
            \Theta\left( 1 \right) & \text{if}~0 < \alpha < 1~\text{(dense)}
            \end{array} \right..
\end{equation}
\end{theorem}
\end{framed}

\begin{IEEEproof}
As a lower-bound we compute the time it takes for the single packet flooding scheme to terminate. Let $K_t$ denote the number of nodes that have the packet after $t$ slots; where $K_1 = 1$. Let $T_N$ be the flooding time, i.e., the first time when $K_t = N$. Let $A_i$, for $1 \leq i \leq K_t$, be the number of new nodes to which node $i$ transmits the packet in slot $t+1$. We then have
\begin{equation}
K_{t+1} = K_t + \sum_{i=1}^{K_t}A_i.
\end{equation}
Since $\expect{A_i | K_t } \leq (N-1)a_N$, we have
\begin{align}
\expect{ K_{t+1} | K_t } &= \expect{K_t + \sum_{i=1}^{K_t}A_i | K_t}, \\
&\leq K_t \left( 1 + (N-1) a_N \right),
\end{align}
for all $t \geq 1$. Applying this recursively, we obtain
\begin{equation}
\label{delay_lb:eq1}
\mathbb{E}\left[ K_t\right] \leq \left( 1 + (N-1) a_N \right)^t.
\end{equation}
Now, using Markov inequality we have
\begin{align}
\mathbb{E}\left[ T_N \right] &\geq t\mathbb{P}\left[ T_N > t\right].
\end{align}
The event $\{ T_N > t\}$ is same as $\{ K_t < N\}$. Hence, we have
\begin{align}
\expect{ T_N } &\geq t\mathbb{P}\left[ K_t < N\right], \\
&= t\left( 1 - \prob{K_t \geq N}\right), \\
&\geq t \left( 1 - \frac{\expect{K_t}}{N} \right),
\end{align}
where the last inequality follows from Markov inequality. Using~\eqref{delay_lb:eq1}, we obtain
\begin{equation}
\label{eq:local1}
\expect{T_N} \geq t \left( 1 -  \frac{1}{N} \left( 1 + (N-1) a_N \right)^t \right),
\end{equation}
for all $t \geq 1$.
Since~\eqref{eq:local1} is a valid lower-bound for all values of $t \geq 1$, setting $t = \frac{1/2 \log N}{\log \left( 1 + (N-1)a_N \right)}$ for $\alpha \geq 1$ and $t = \frac{1/2 \log N^{\alpha}}{\log \left( 1 + (N-1)a_N \right) }$ for $0 < \alpha < 1$ yields the result.

\end{IEEEproof}

\begin{figure}
\centering
\includegraphics[width=0.85\linewidth]{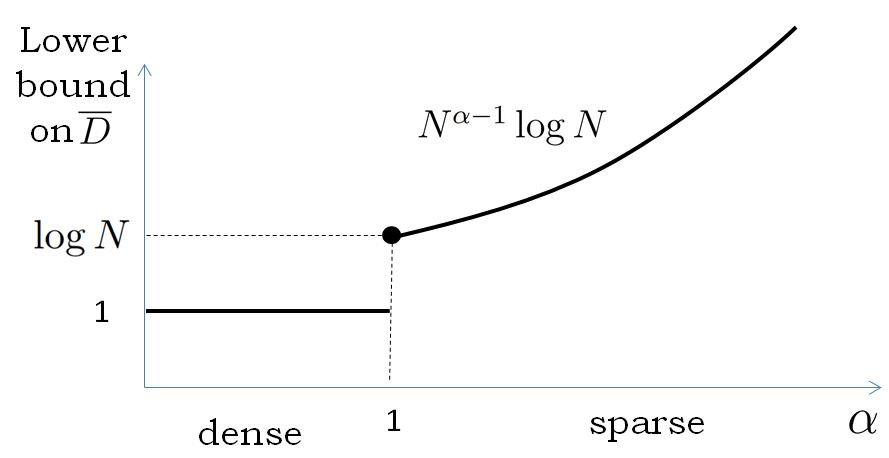}
\caption{Lower bound on achievable average delay $\overline{D}$ as a function of $\alpha$.}
\label{fig:min_delay}
\end{figure}

In Figure~\ref{fig:min_delay}, we plot the lower-bound on average delay $\overline{D}$ as a function of $\alpha$. We observe that as the network gets sparser the number of nodes receiving the flooded packet per cell decreases, thereby, increasing the broadcast delay. Thus, the lower-bound is a non-decreasing function of $\alpha$. However, for $0 < \alpha < 1$ the delay bound is a constant $O(1)$, and remains unchanged. Clearly, if $C = 1$, i.e. if the entire network is a single cell, then the broadcast delay will be $1$ as the packet can reach all other nodes in a single transmission.
In the next two sections we show that this lower-bound on average delay is in fact achievable, up to $\log \log N$ factor.

\section{Flooding Time in Markov Evolving Graphs}
\label{sec:meg}
In order to gains further insights into the flooding time of the packet flooding scheme we use the theory of Markov evolving graphs (MEG), to help us derive the necessary upper bound on the flooding time. We start with a brief introduction to MEG and a review of pertinent results.

Let $\mathcal{G}$ be a family of graphs with node set $[N] = \left\{ 1, 2, \ldots N\right\}$. The Markov chain $\mathcal{M} = \left( G_t\right)_{t \in \mathbb{N}}$, where $G_t \in \mathcal{G}$, with state space $\mathcal{G}$ is called a MEG. Note that $\mathcal{G}$ is a finite set. For our network model of Figure~\ref{fig:sys_model}, if we draw edge between $i$ and $j$ whenever both nodes $i$ and $j$ lie in the same cell, the resulting time evolving graph is an MEG. When the MEG has a unique stationary distribution we call it a stationary MEG.\footnote{Since the state space $\mathcal{G}$ is finite, it always has at least one stationary distribution.} In this work, we assume that a stationary MEG starts from it's stationary distribution. The IID mobility model results in one such  stationary MEG, as every graph formation can follow any other in $\mathcal{G}$. We now describe the single packet flooding scheme in MEG.

\begin{framed}
\underline{Single packet flooding for a MEG}: In the first slot only a single node $s$ has the packet, i.e.\ $I_1 = \{s\}$. Here, $I_t \subset [N]$ denotes the set of nodes that have the packet at time $t$. In every slot $t \geq 1$:
\begin{enumerate}
  \item Identify the neighbors of $I_t$ that are not in $I_t$:
  \begin{equation}
  N(I_t) = \left\{ \text{neighbours of}~I_t~\text{in}~G_t\backslash I_t \right\}.
  \end{equation}

  \item Transmit the packet to each node in $N(I_t)$. We, thus, have
  \begin{equation}
  I_{t+1} = I_t \bigcup N(I_t).
  \end{equation}

  \item If $I_t = [N]$ then stop, else start again from Step~1.
\end{enumerate}
\end{framed}
Let $T_N$ be the \emph{flooding time}, i.e., the time it takes for this process to terminate. Note that, this scheme reduces to the single packet flooding scheme of~Section~\ref{sec:cap_delay} for our network model.
An upper bound on flooding time was derived in~\cite{clementi11_MEG}. This bound depended on the MEG satisfying certain expander properties. We summarize this result in Theorem~\ref{thm:MEG_expander}, and provide two new bounds on flooding time in Theorem~\ref{thm:MEG_geobound} and Theorem~\ref{thm:MEG_hybrid}.

The expander property of MEG is defined in terms of the expander property of a static graph~\cite{clementi11_MEG}.
\begin{framed}
\begin{definition}
A graph $G = \left([N], E\right)$ is said to be $([h_0, h_1], k)$-expander if for every $I \subset [N]$ such that $h_0 < |I| \leq h_1$ we have
\begin{equation}
|N(I)| \geq k|I|,
\end{equation}
where $N(I)$ is the set of all neighbours of nodes in $I$ that are not already in $I$.
\end{definition}
\end{framed}
We now use this to define the expander property of MEG.
\begin{framed}
\begin{definition}
Stationary MEG $\mathcal{M} = \left( G_t \right)_{t \in \mathbb{N}}$ is $\left( [h_0, h_1], k\right)$-expander with probability $p$ if
\begin{equation}
\prob{ G_0~\text{is}~\left( [h_0, h_1], k\right)\text{-expander} } \geq p.
\end{equation}
\end{definition}
\end{framed}
If the graph is $([h-1,h],k)$-expander then for notational simplicity we say that it is $(h,k)$-expander. To show that a stationary MEG is $(h,k)$-expander we have to evaluate the probability
\begin{equation}
\prob{ \bigcap_{|I| = h} \left\{ |N(I)| \geq k|I|\right\} }.
\end{equation}
%Also, instead of saying that the graph is $([h_i, h_{i+1}], k_i)$-expander we say that the graph is $(h,k_i)$-expander for all $h_i < h \leq h_{i+1}$.
The following upper bound on flooding time was derived in~\cite{clementi11_MEG}.
\begin{framed}
\begin{theorem}{\cite{clementi11_MEG}}
\label{thm:MEG_expander}
For a stationary MEG, if
\begin{equation}
\label{eq:tic1}
\prob{ \bigcap_{i=1}^{s}\left\{ G_0~\text{is an}~\left([h_{i-1}, h_i], k_i \right)\text{-expander} \right\} } \geq_N 1 - \frac{c_1}{N^2}
\end{equation}
for some $c_1>0$, $1 = h_0 \leq h_1 < h_2 < \cdots < h_s = \frac{N}{2}$, a non-increasing sequence $k_1 \geq k_2 \geq \cdots \geq k_{s} > 0$, and $s \in \{2, 3, \ldots \frac{N}{2}\}$ then the flooding time
\begin{equation}
T_N = O\left( \sum_{i = 1}^{s} \frac{\log\left( h_i / h_{i-1} \right)}{\log(1 + k_{i})}\right),
\end{equation}
with probability at least $1 - \frac{c_2}{N}$ for some $c_2 >0$.
\end{theorem}
\end{framed}

A stationary MEG may not always satisfy the expander property required by~\eqref{eq:tic1}. In such a case, we provide the following two bounds for flooding time for a stationary MEG.
\begin{framed}
\begin{theorem}
\label{thm:MEG_geobound}
If for every $h \in {1, 2, \ldots N-1}$ and for all $I \subset [N]$ with $|I| = h$, there exists a function $p(h)$ such that $\prob{N(I) = 1} \geq_N p(h) > 0$ then the flooding time
\begin{equation}
T_N = O\left( \sum_{h=1}^{N - 1} \frac{1}{p(h)}\right),
\end{equation}
with probability at least $1 - e^{-c_1 N}$ for some $c_1 > 0$.
\end{theorem}
\end{framed}

%{\em Proof}:
\begin{IEEEproof}
We denote $X \sim \text{Geo}(p)$ when $X$ is a geometrically distributed random variable with parameter $p$, that is, $\prob{X=k} = p \left( 1 - p\right)^{k-1}$ for all $k \geq 1$. Let $X_h \sim \text{Geo}\left( \prob{N(h) = 1} \right)$ and $Z_h \sim \text{Geo}\left( p(h) \right)$ for all $h \in \{1, 2, \ldots N-1\}$. It is clear that $X_h \leq_N Z_h$ a.s. for all $1 \leq h \leq N-1$. If the packet transmissions were to take place only at the occurrences of the events $\{ N(h) = 1\}$, the flooding time would be much larger, and would equal $\sum_{h=1}^{N-1} X_h$. This implies
\begin{equation}
T_N \leq \sum_{h=1}^{N-1} X_h
\end{equation}
Further, since $\prob{N(h) = 1}  \geq_N p(h)$ we have $X_h \leq_N Z_h$ a.s. for all $h$. This implies
\begin{equation}
T_N \leq \sum_{h=1}^{N-1} X_h \leq_N \sum_{h=1}^{N-1} Z_h. \label{eq:localx}
\end{equation}
Now, using the concentration bound given in Lemma~\ref{lem:geo_cbound} of Appendix~\ref{pf:concentration_bounds} on $\{Z_1, \ldots Z_{N-1}\}$ and substituting $t = \mu =  \sum_{h=1}^{N - 1} \frac{1}{p(h)}$ we obtain
\begin{equation}
\prob{ \sum_{h=1}^{N-1} Z_h > 2c_1\mu } \leq \left( 1 - p^{\ast} \right)^{\mu}\exp\left\{ -\frac{2c_1-3}{4}(N-1)\right\},
\end{equation}
for some $c_1 \geq 2$, where $p^{\ast} = \min_{h \in \{1, 2, \ldots N-1\}} p(h)$. Note that $\left(1-p^{\ast}\right)^{\mu} \leq 1$. We, thus, have
\begin{align}
\prob{ \sum_{h=1}^{N-1} Z_h > 2c_1\mu } &\leq  \exp\left\{ -\frac{2c_1 -3}{4}(N-1)\right\} \\ &= \Theta\left( \exp\{ -c_2 N\}\right), \label{eq:localy}
\end{align}
for some positive constant $c_2$.
From~\eqref{eq:localx} and~\eqref{eq:localy} we have
\begin{equation}
\prob{T_N \leq 2c_1 \sum_{h=1}^{N - 1} \frac{1}{p(h)} } \geq_N 1 -  \exp\{ -c_2 N\}.
\end{equation}
%This completes the proof.~\qed
\end{IEEEproof}

Notice that instead of $\prob{N(I) = 1} \geq_N p(h) > 0$ if we have the condition $\prob{N(I) \geq 1} \geq_N p(h) > 0$ the same result holds, using an identical proof.

Theorem~\ref{thm:MEG_geobound}, does not use any expander properties of the MEG. It can happen that a stationary MEG satisfies the expander property for some subsets $I \subset [N]$ but not all. In this case Theorem~\ref{thm:MEG_geobound} may not give a very tight bound. We can combine the ideas of Theorem~\ref{thm:MEG_expander} and~\ref{thm:MEG_geobound} to establish the following result.
\begin{framed}
\begin{theorem}
\label{thm:MEG_hybrid}
For a stationary MEG if
\begin{enumerate}
  \item there exists a $s \in \{2, 3, \ldots \frac{N}{2}\}$, strictly increasing sequence $1 < h_1 < h_2 < \cdots < h_s = \frac{N}{2}$, and a non-increasing sequence $k_2 \geq k_3 \geq \cdots \geq k_s > 0$ such that
      \begin{multline}
      \label{eq:qu1}
      \prob{ \bigcap_{i=2}^{s}\left\{ G_0~\text{is}~\left([h_{i-1}, h_i], k_i \right)\text{-expander} \right\} } \\
      \geq_N 1 - \frac{c_1}{N^2},
      \end{multline}
      for some $c_1 > 0$,

  \item for $1 \leq h \leq h_1$, for all $I \subset [N]$ such that $|I| = h$ we have
      \begin{equation}
      \label{eq:qu2}
      \prob{N(I) = 1} \geq_N p(h) > 0,
      \end{equation}
      and

  \item $h_1 \geq c_2 \log N$ is such that
      \begin{equation}
      \label{eq:qu3}
      \lim_{N \rightarrow \infty} \frac{h_1}{\log N} = \infty,
      \end{equation}
\end{enumerate}
then
\begin{equation}
T_N = O\left( \sum_{h=1}^{h_1} \frac{1}{p(h)} + \sum_{i=2}^{s} \frac{\log\left( h_{i}/h_{i-1} \right)}{\log\left(1 + k_i\right)} \right),
\end{equation}
with probability at least $1 - c_2/N$ for some $c_2 > 0$.
\end{theorem}
\end{framed}

% {\em Proof}:
\begin{IEEEproof}
$I_t \subset [N]$ denotes the number of nodes that have the packet at time $t \geq 1$. Let $T_1$ be the first time at which at least $h_1$ nodes get the packet, i.e.,
\begin{equation}
T_1 = \min\left\{ t \geq 1 | |I_t| \geq h_1~\text{and}~|I_1| = 1\right\},
\end{equation}
and $T_{2:N} = T_N - T_1$. Clearly, $T_{2:N}$ will be less than the time it takes for the packet to reach all nodes if the system were to start with exactly $h_1$ nodes carrying the packet, i.e.,
\begin{equation}
T_{2:N} \leq T_{2:N}^{'} = \min\left\{  t \geq 1 | |I_t| = N~\text{and}~|I_1| = h_1 \right\}.
\end{equation}
Following the same arguments listed in~\cite{clementi11_MEG} for the proof of Theorem~\ref{thm:MEG_expander}, while using the expander property~\eqref{eq:qu1}, we have
\begin{equation}
\label{eq:nap1}
T_{2:N}^{'} = O\left(  \sum_{i=2}^{s} \frac{\log\left( h_{i}/h_{i-1} \right)}{\log\left(1 + k_i\right)} \right),
\end{equation}
with probability at least $1 - c_1/N$ for some $c_1 > 0$.

Following the same arguments in the proof of Theorem~\ref{thm:MEG_geobound}, while using~\eqref{eq:qu2}, yields
\begin{equation}
T_1 = O\left(  \sum_{h=1}^{h_1} \frac{1}{p(h)} \right),
\end{equation}
with probability at least $1 - \exp\left\{ -c_2 h_1\right\}$ for some $c_2 > 0$. From~\eqref{eq:qu3}, it is clear that $h_1 > \gamma\log N$ for any $\gamma > 0$. This implies
\begin{align}
1 - \exp\left\{ -c_2 h_1\right\} &\geq 1 - \exp\left\{ -c_2 \gamma \log N \right\}, \\
&\geq 1 - \frac{1}{N^{c_2 \gamma}},
\end{align}
for any $\gamma > 0$. Choosing any $\gamma \geq 1/c_2$ yields
\begin{equation}
\label{eq:nap2}
T_1 = O\left(  \sum_{h=1}^{h_1} \frac{1}{p(h)} \right),
\end{equation}
with probability at least $1 - c_3/N$ for some $c_3 > 0$. We know that $T_N \leq T_1 + T_{2:N}^{'}$. Using~\eqref{eq:nap1} and~\eqref{eq:nap2} we obtain the desired result.%~\qed
\end{IEEEproof}

The results also hold if we replace the condition $\prob{N(I) = 1} \geq_N p(h) > 0$ with
\begin{equation}
\prob{N(I) \geq 1} \geq_N p(h) > 0.
\end{equation}

Theorems~\ref{thm:MEG_expander},~\ref{thm:MEG_geobound}, and~\ref{thm:MEG_hybrid} give a high probability upper bound on flooding time, and not an upper bound on average flooding time. In the next section we apply these results to obtain a high probability upper bound on flooding time for our network model, and show that it nearly scales as the lower bound on average flooding time obtained in Theorem~\ref{thm:delay_lb} of Section~\ref{sec:cap_delay}. In Section~\ref{sec:fcfs_flood}, we use this fact to propose a FCFS flooding scheme that achieves the high probability upper bound as its average delay.

\section{Flooding Time for the IID Mobility Model}
\label{sec:flood_time}
We now apply the high probability upper bounds on flooding time from Theorems~\ref{thm:MEG_expander},~\ref{thm:MEG_geobound}, and~\ref{thm:MEG_hybrid} of Section~\ref{sec:meg} to our network model. As to which of the three results we use depends on whether the network is sparse or dense. Let $\mathcal{M}$ denote the stationary MEG for our network model of Figure~\ref{fig:sys_model}, and let $G_0$ be it's stationary distribution.
\begin{framed}
\begin{theorem}
\label{thm:flooding_time}
The flooding time is
\begin{equation}
T_N = \left\{ \begin{array}{ll}
            O\left( N^{\alpha - 1} \log N\right) & \text{if}~\alpha \geq 1~\text{(sparse)}\\
            O\left( \log \log N \right) & \text{if}~0 < \alpha < 1~\text{(dense)}
            \end{array} \right.,
\end{equation}
with probability at least $1 - \frac{c_1}{N}$ for some $c_1 > 0$.
\end{theorem}
\end{framed}

%{\em Proof}:
\begin{IEEEproof}
We derive this by showing the expander properties of the network $\mathcal{M}$. We split the proof into three cases: $0 < \alpha < 1$, $1 \leq \alpha < 2$, and $\alpha \geq 2$.
\begin{enumerate}
  \item \underline{$0 < \alpha < 1$}: In this case, the expander properties of Theorem~\ref{thm:MEG_expander} hold. Note that
  \begin{equation}
  \expect{N(h)} = (N-h)\left[ 1 - \left( 1 - c/N^\alpha \right)^h\right]. \label{eq:localm}
  \end{equation}
  It is also easy to see that $1 - \left( 1 - c/N^\alpha \right)^h = \Theta\left(h/N^\alpha\right)$ if $h/N^\alpha \rightarrow 0$, and $1 - \left( 1 - c/N^\alpha \right)^h = \Theta(1)$ if $h/N^\alpha \rightarrow \infty$. When $h/N^\alpha = \Theta(1)$, both are true. We, therefore, have
      \begin{equation}
      \expect{N(h)} = \left\{ \begin{array}{ll}
            \Theta\left( N h/N^\alpha \right) & \text{for}~1 \leq h \leq N^{\alpha} \\
            \Theta(N) &\!\!\!\!\!\!\!\!\!\!\!\! \text{for}~N^{\alpha} + 1 \leq h \leq N/2
            \end{array} \right..
      \end{equation}
      Since, in both cases we have $\expect{N(h)} \rightarrow \infty$, we can use Lemma~\ref{lem:binom_2}, the concentration bound on the binomial distribution, to show that the event $\{N(h) \geq c_1 \expect{N(h)}\}$ occurs with high probability for some $0 < c_1 < 1$. This proves that the graph is $(h, k(h))$-expander where $k(h) = c_1 \frac{\expect{N(h)}}{h}$, i.e.,
  \begin{equation}
  \label{eq:expander_large}
  \prob{  \bigcap_{h = 2}^{N/2} \left\{ G_0~\text{is}~(h,k(h))\text{-expander} \right\} } \geq_N 1 - \frac{c_2}{N^2},
  \end{equation}
  for some $c_2 > 0$ where
  \begin{equation}
  k(h) = \left\{ \begin{array}{ll}
            c_3 N^{1-\alpha} &\text{for}~1 \leq h \leq N^{\alpha} \\
            c_4 \frac{N}{h} & \text{for}~N^{\alpha} + 1 \leq h \leq N/2
            \end{array} \right.,
  \end{equation}
  for some $c_3, c_4 > 0$. See Appendix~\ref{pf:expander_large} for a detailed proof. This satisfies the expander property requirements of Theorem~\ref{thm:MEG_expander}. Applying Theorem~\ref{thm:MEG_expander}, we obtain
  \begin{equation}
  T_N = O\left( \log \log N\right),
  \end{equation}
  with probability at least $1 - \frac{c_5}{N}$ for some $c_5 > 0$. We prove this in Appendix~\ref{pf:expander_large}.

  \item \underline{$1 \leq \alpha < 2$}: In this case, the expander properties of Theorem~\ref{thm:MEG_hybrid} hold. Note that $\frac{h}{N^\alpha} \rightarrow 0$ for all $1 \leq h \leq N/2$. We, thus, have $\left( 1- \left( 1 - c/N^\alpha\right)^h\right) = \Theta\left( h/N^\alpha\right)$. Using the expression for $\expect{N(h)}$ in~\eqref{eq:localm} we have $N(h) = \Theta\left( N h/N^\alpha\right) = \Theta\left( h/N^{\alpha - 1}\right)$.

      Here, $\expect{N(h)}$ does not always go infinity in $N$. However, we observe that, for all $\beta N^{\alpha - 1}\log N + 1 \leq h \leq N/2$ and for any $\beta > 0$, $\expect{N(h)} \rightarrow \infty$ as $N \rightarrow \infty$. We can then use Lemma~\ref{lem:binom_2}, the concentration bounds for binomial distribution, to derive the following expander property for $\beta N^{\alpha - 1}\log N + 1 \leq h \leq N/2$:
      \begin{align}
      \label{eq:expander_small1}
      &\prob{ \bigcap_{h > \beta N^{\alpha-1}\! \log N }^{N/2}\!\!\left\{ G_0~\text{is}~\!\left(h,\frac{c_1}{N^{\alpha-1}}\right)\text{-expander} \right\} \!} \nonumber \\ &\geq_N 1 - \frac{c_2}{N^2},
      \end{align}
      for some $c_1, c_2 > 0$ and provided $\beta > c_3$ for some $c_3 > 0$.

      For $1 \leq h \leq \beta N^{\alpha - 1}\log N$, $\expect{N(h)}$ need not always go to infinity, and can in fact go to zero. Due to this, the network $\mathcal{M}$ does not satisfy any expander property for all $1 \leq h \leq \beta N^{\alpha - 1}\log N$. Therefore, we derive a lower-bound on the probability $\prob{N(h) \geq 1}$. In particular, there exists $c_3 > 0$ such that
      \begin{equation}
      \label{eq:expander_small2}
      \prob{N(h) \geq 1} \geq_N c_3 \left( 1 - \exp\left\{ - h/N^{\alpha - 1} \right\} \right),
      \end{equation}
      for all $h \in \{1, 2, \ldots \beta N^{\alpha-1} \log N \}$. See Appendix~\ref{pf:expander_small} for a detailed proof. This satisfies the conditions of Theorem~\ref{thm:MEG_hybrid}. From this, one can obtain
      \begin{equation}
      T_N = O\left( N^{\alpha - 1} \log N\right), \nonumber
      \end{equation}
      with probability at least $1 - \frac{c_4}{N}$ for some $c_4 > 0$. We prove this in Appendix~\ref{pf:expander_small}.

  \item \underline{$\alpha \geq 2$}: In this case, the conditions of Theorem~\ref{thm:MEG_geobound} hold. Since $\alpha \geq 2$, we have $h/N^\alpha \rightarrow 0$ for all $1 \leq h \leq N/2$. This implies $1 - \left( 1 - c/N^\alpha\right)^h  = \Theta\left(h/N^\alpha\right)$. Thus, using~\eqref{eq:localm}, we have $\expect{N(h)} = \Theta\left( Nh/N^\alpha\right) \rightarrow 0$ for all $1 \leq h \leq N/2$. This shows that the network $\mathcal{M}$ does not satisfy any expander property. We, therefore, derive a lower-bound on $\prob{N(h) = 1}$.
  There exists a $c_1 > 0$ such that
  \begin{equation}
  \label{eq:expander_tiny}
  \prob{N(h) = 1} \geq_N c_1 \frac{(N-h)h }{N^\alpha},
  \end{equation}
  for all $1 \leq h \leq N-1$. See Appendix~\ref{pf:expander_tiny} for a detailed proof. This satisfies the condition of Theorem~\ref{thm:MEG_geobound}, using which one can obtain
  \begin{equation}
  T_N = O\left( N^{\alpha - 1} \log N\right), \nonumber
  \end{equation}
  with probability at least $1 - \frac{c_2}{N}$ for some $c_2 > 0$. We prove this in Appendix~\ref{pf:expander_tiny}.
\end{enumerate}
% Detailed arguments are given in the technical report~\cite{tech_report}.~\qed
\end{IEEEproof}

\begin{figure}
\centering
\includegraphics[width = 0.85\linewidth]{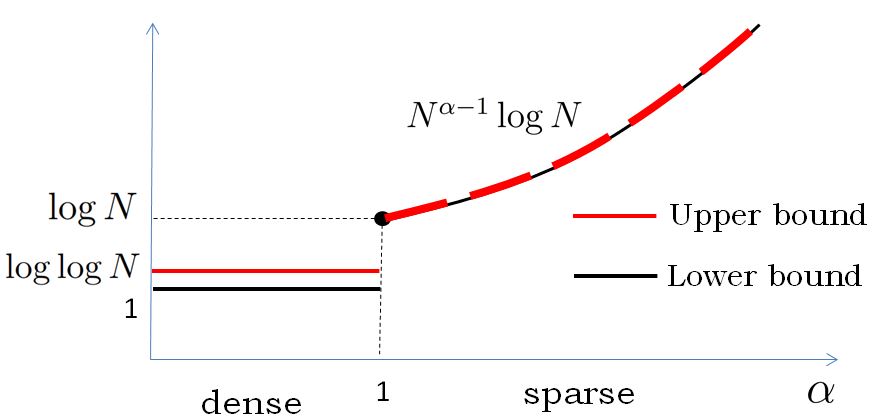}
\caption{High probability upper bound and the average lower-bound on flooding time $T_N$ as a function of $\alpha$.}
\label{fig:delay_compare}
\end{figure}

Figure~\ref{fig:delay_compare} compares the high probability upper bound with the average lower-bound on flooding time $T_N$ from Theorem~\ref{thm:delay_lb}. We observe a gap of at most $O\left(\log \log N\right)$ when $0 < \alpha < 1$. For all other values of $\alpha$ the upper and lower-bounds are of the same order. The lower-bound on flooding time was derived in Theorem~\ref{thm:delay_lb}, which was also the lower-bound on the achievable average delay. In the next section, we show that a simple FCFS flooding scheme achieves the high probability upper bound on flooding time as its achievable average delay.

\section{FCFS Flooding Scheme}
\label{sec:fcfs_flood}
We propose a scheme that is based on the idea of single packet flooding described in Section~\ref{sec:cap_delay}. In this scheme, only a single packet is transmitted over the entire network at any given time. Packets are served sequentially by the network on a FCFS basis. Each packet gets served for a fixed duration of $U_N$. The packet is dropped if within this duration it is not received by all the other $(N-1)$ nodes. We call this the \emph{FCFS packet flooding} scheme.

\begin{framed}
\underline{FCFS Packet Flooding}: Packets arrive at each of the $N$ nodes at rate $\lambda$.
\begin{enumerate}
  \item Among all the packets that have arrived, select the one that had arrived the earliest. At this time only one node, i.e.\ the source node, has this packet.

  \item In every cell, randomly select one packet carrying node (if it exists) as a transmitter.

  \item Selected nodes transmit in each cell during the slot while all other nodes in the corresponding cells receive the packet.

  \item Repeat Steps~2 and~3 for $U_N$ time slots.

  \item After $U_N$ slots, remove the current packet from the transmission queue and go to Step~1.
\end{enumerate}
\end{framed}

Since we abruptly terminate the process in Step~5 after $U_N$ slots, it can happen that the packet has not reached all the $(N-1)$ destination nodes. To ensure that this happens rarely let
\begin{equation}
\label{eq:un}
U_N = \left\{ \begin{array}{ll}
            c_1 N^{\alpha - 1} \log N  & \text{if}~\alpha \geq 1~\text{(sparse)}\\
%            c_2 \log N  & \text{if}~\alpha = 1~\text{(critical)} \\
            c_2 \log \log N  & \text{if}~0 < \alpha < 1~\text{(dense)}
            \end{array} \right.,
\end{equation}
for some positive constants $c_1$ and $c_2$ such that $T_N < U_N$ with probability $1 - \frac{1}{N}$. Such constants exists by Theorem~\ref{thm:flooding_time}. This leads to a vanishingly small packet drop rates. We now obtain the capacity and delay performance of this FCFS packet flooding scheme.

\begin{framed}
\begin{theorem}
\label{thm:fcfs_pkt_flooding}
The FCFS packet flooding scheme achieves a capacity of
\begin{align}
\label{eq:fcfs_flooding_cap}
\lambda = \left\{ \begin{array}{ll}
            \Theta\left( \frac{1}{N^{\alpha} \log N}\right) & \text{if}~\alpha \geq 1~\text{(sparse)} \\
            % \Theta\left( \frac{1}{N \log N}  \right)   & \text{if}~\alpha = 1~\text{(critical)} \\
            \Theta\left( \frac{1}{N \log \log N}\right) & \text{if}~0 < \alpha < 1~\text{(dense)}
            \end{array} \right..
\end{align}
Furthermore, the delay achieved at this rate is $\overline{D} = \Theta\left(U_{N}\right)$.
\end{theorem}
\end{framed}
\begin{IEEEproof}
The packets arrive at each node according to a Poisson process, at rate $\lambda$. Thus, the sum packets arrivals in the networks is also a Poisson process of rate $N\lambda$. The service time for each packet under the FCFS packet flooding scheme is nothing but $U_N$. Thus, the system can be thought of as a M/D/1 queue, with an arrival rate of $N\lambda$ and service time of $U_N$. The waiting time for such a system is given by~\cite{data_nets}
\begin{equation}
\label{eq:fcfs_flooding_delay}
\tilde{W} = U_N + U_N \frac{\rho}{2(1-\rho)},
\end{equation}
for any arrival rate $N\lambda < \frac{1}{U_N}$, where $\rho = N U_{N} \lambda < 1$ is the queue utilization. Selecting any $\rho < 1$, we obtain $\tilde{W} = \Theta(U_N)$ and $\lambda = \Theta\left( \frac{1}{N U_N}\right)$. Substituting $U_N$ from~\eqref{eq:un}, we obtain the result.
\end{IEEEproof}

This implies that the delay lower-bound of Theorem~\ref{thm:delay_lb} is achieved, up to a gap of $O\left( \log \log N\right)$, when the network is dense, i.e.\ $0 < \alpha < 1$.
We also see that the achieved throughput $\lambda$ is less than the capacity upper bound of Theorem~\ref{thm:capacity} by a factor of $\log \log N$ when $0 < \alpha < 1$, and by a factor of $\log N$, when $\alpha \geq 1$. The $\log \log N$ gap appears due to the exact same gap between the flooding time upper and lower bounds when $0 < \alpha < 1$. The $\log N$ factor gap for $\alpha \geq 1$ occurs even though the flooding time upper and lower bounds are asymptotically tight. This, we conjuncture, is because the FCFS flooding scheme does not allow simultaneous transmissions of different packets, which leads to inefficient utilization of available transmission opportunities.
%
%%% \rt{Reasons for the gap}.

We summarize these results in Table~\ref{tbl:rate}. Unlike the unicast case, where a capacity-delay tradeoff has been observed~\cite{modiano, dev_shah06, sharma_ravi05}, nearly no such tradeoff exists for the broadcast problem, and both capacity and minimum delay can be nearly achieved simultaneously.

%\begin{figure}
%\centering
%\includegraphics[width = 0.85\linewidth]{tradeoff2}
%\caption{Achievable delay versus throughput plot. The performance of FCFS flooding scheme is represented by a black dot.}
%\label{fig:tradeoff}
%\end{figure}

\section{Single Hop Scheme}
\label{sec:single_hop}
We now propose a single-hop scheme that achieves the capacity upper-bound of Theorem~\ref{thm:capacity} when the network is sparse, i.e.\ $\alpha \geq 1$. In this scheme, a packet reaches it's destination from a source in a single hop, i.e.\ by direct source to destination transmission. This scheme only allows for a single receiver in each cell, thus, ignores the broadcast nature of the wireless medium. The scheme still achieves the upper-bound capacity as the number of nodes in a cell tends to be very small in the sparse case.

\begin{figure}
\centering
\includegraphics[width = 0.75\linewidth]{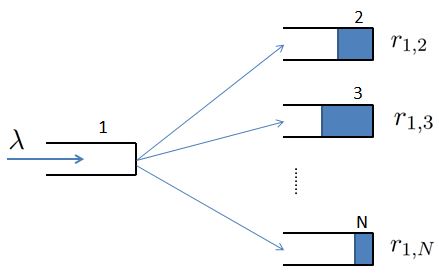}
\caption{Node~1 makes $(N-1)$ copies of every arriving packet, one for each queue $Q_{1,j}$ for $2 \leq j \leq N$. Service rate of $Q_{1,j}$ is denoted by $r_{1,j}$.}
\label{fig:single_hop}
\end{figure}

\begin{framed}
\underline{Single-Hop Scheme}: Each node makes $(N-1)$ copies of an arrival packet, one for each receiving node. Figure~\ref{fig:single_hop} illustrates this for node $1$, where a copy of an arriving packet at node $1$ is transferred to each of the queues $Q_{1,j}$ for all $2 \leq j \leq N$.
\begin{enumerate}
\item In each cell, select a pair of nodes at random. If a cell contains fewer than $2$ nodes no transmissions occur in that cell.

\item For the selected pair in every cell, assign, uniformly and randomly, one node as a transmitter and the other as receiver.

\item For each transmitter-receiver pair, if the transmitter node has a packet for the receiver node, transmit it, else remain idle.

\item Wait for the next slot to begin, and restart the process from Step~1.
\end{enumerate}
\end{framed}

The scheme is opaque to which node pairs are chosen as the source-destination pairs. Thus, every queue $Q_{i,j}$ is activated at the same rate. This implies that all the queues $Q_{i,j}$ have identical service rates. Hence,
\begin{equation}
\label{eq:amm}
\sum_{i \neq j} r_{i,j} = N(N-1) r_{1,2}.
\end{equation}
The left hand side of~\eqref{eq:amm} corresponds to the total tate of service opportunities across the network, which is given by $Cp$, where $p$ is the probability that there are at least two nodes in a cell: $p = 1 - \left( 1 - a_N \right)^N - Na_N\left( 1 - a_N\right)^{N-1}$. Thus, $N(N-1) r_{1,2} = Cp$, which gives,
\begin{equation}
r_{1,2} = \frac{Cp}{N(N-1)}.
\end{equation}
Hence, any arrival rate $\lambda < r_{1,2}$ will yield a stable network under the single-hop scheme. The delay achieved by this scheme is lower-bounded by the delay in the single queue. Since each queue is Bernoulli arrival and Bernoulli service, the waiting time in each queue is given by $\bar{W} = \frac{1 - \lambda}{r_{1,2} - \lambda}$. Setting $\lambda = \frac{1}{2} r_{1,2}$ we obtain $\bar{W} = \Theta\left( 1/ r_{1,2} \right)$. We summarize this in the following result.
\begin{framed}
\begin{theorem}
\label{thm:single_hop}
The single hop scheme achieves a capacity of
\begin{equation}
\lambda_{\text{SH}}= \left\{ \begin{array}{ll}
            \Theta\left( \frac{1}{N^{\alpha}}\right) & \text{if}~\alpha \geq 1~\text{(sparse)} \\
            % \Theta\left(\frac{1}{N}\right) & \text{if}~\alpha = 1~\text{(critical)} \\
            \Theta\left( \frac{1}{N^{2-\alpha} }\right) & \text{if}~0<\alpha<1~\text{(dense)}
            \end{array} \right.,
\end{equation}
Furthermore, the delay achieved at this rate is
\begin{equation}
\overline{D}_{\text{SH}} \geq \left\{ \begin{array}{ll}
            \Theta\left( N^{\alpha} \right) & \text{if}~\alpha \geq 1~\text{(sparse)} \\
            % \Theta\left( N\right) & \text{if}~\alpha = 1~\text{(critical)} \\
            \Theta\left( N^{2-\alpha} \right) & \text{if}~0<\alpha<1~\text{(dense)}
            \end{array} \right..
\end{equation}
\end{theorem}
\end{framed}
Hence, the single hop scheme achieves the capacity upper-bound for $\alpha \geq 1$. Thus, the capacity upper bound in Theorem~\ref{thm:capacity} is indeed achievable.

\section{Conclusion}
\label{sec:conclusion}
We considered the problem of all-to-all broadcast transmissions, in a networks of highly mobile nodes.
We derived the broadcast capacity and minimum delay scaling, in the number of vehicles $N$, and showed that the capacity cannot scale better than $1/N$. This, in conjunction with earlier known results for static network~\cite{taliv2006}, proves that the broadcast capacity does not improve with high mobility. This is in contrast with the unicast case for which mobility improves network capacity~\cite{tse02}.

We further showed that both, the capacity and minimum delay scalings, can be nearly achieved, simultaneously. We proposed a simple FCFS flooding scheme, that nearly achieves this both capacity and minimum delay scaling. The flooding time bound for Markov evolving graphs (MEG), proposed in~\cite{clementi11_MEG}, was used to analyze the proposed scheme. We derived two new bounds on flooding time for MEG, which may be of independent theoretical interest.

%%%%%%%%%%%%%%%%%%%%%%%%%%%%%%%%%%%%%%%%%%%%%%%%%%%%%%%%%%%%%%%%%%%%%%%%%%%%%%%%%%%%%%%%%%%%%%%%%%%%%%%%%%%%%%%%%%%
%
\appendix

\subsection{Proof of Theorem~\ref{thm:capacity}}
\label{pf:thm:capacity}
Let $\lambda$ be the rate achieved by a scheme. If $X_{h}(T)$ is the number of packets delivered to the destination in exactly $h$ hops by time $T$ then for an $\epsilon > 0$ we have
\begin{equation}
\label{thput_varcell_thm_proof:eq1}
\frac{1}{T}\sum_{h \geq 1} X_{h}(T) > N(N-1)\lambda - \epsilon
\end{equation}
for all $T > T_{\epsilon}$, for some $T_{\epsilon} > 0$.

If $Z_{i}^{k}(t)$ is a binary random variable which equals $1$ if there are $k$ nodes in cell $i$ in slot $t$ then the total number of packet receptions by time $T$ is at most $\sum_{i=1}^{C}\sum_{k = 2}^{N}\sum_{t=1}^{T} (k-1)Z_{i}^{k}(t)$. Hence,
\begin{equation}
\label{thput_varcell_thm_proof:eq2}
\sum_{h \geq 1} hX_{h}(T) \leq \sum_{i=1}^{C}\sum_{k = 2}^{N}\sum_{t=1}^{T} (k-1)Z_{i}^{k}(t).
\end{equation}
Combining~\eqref{thput_varcell_thm_proof:eq1} and~\eqref{thput_varcell_thm_proof:eq2} we obtain
\begin{align}
\sum_{i=1}^{C}\sum_{k = 2}^{N} \frac{1}{T}\sum_{t=1}^{T} (k-1)Z_{i}^{k}(t) &\geq \frac{1}{T} \sum_{h \geq 1} hX_{h}(T), \nonumber \\
&= \frac{1}{T}X_{1}(T) + \frac{1}{T}\sum_{h \geq 2} h X_{h}(T), \nonumber \\
&\geq \frac{1}{T}X_{1}(T) + \frac{2}{T}\sum_{h \geq 2} X_{h}(T). \nonumber
\end{align}
Using~\eqref{thput_varcell_thm_proof:eq1} we obtain
\begin{multline}
\sum_{i=1}^{C}\sum_{k = 2}^{N} \frac{1}{T}\sum_{t=1}^{T} (k-1)Z_{i}^{k}(t) \geq \frac{1}{T}X_{1}(T) \\
+ 2 \left( N(N-1)\lambda -\epsilon - \frac{1}{T}X_{1}(T)\right). \nonumber
\end{multline}
Taking $T \rightarrow +\infty$ we have
\begin{align}
\sum_{i=1}^{C}\sum_{k = 2}^{N} (k-1) p(k) &\geq Cp + 2\left( N(N-1)\lambda - \epsilon - Cp\right),\nonumber \\
&= 2N(N-1) - 2\epsilon -Cp,
\end{align}
where $p(k)$ is the probability that there are $k$ nodes in a cell and $p$ is the probability that there are at least two nodes in a cell; we use the fact that $\limsup_{T \rightarrow +\infty} \frac{X_{1}(T)}{T} \leq Cp$. Taking $\epsilon \rightarrow 0$, we obtain
\begin{align}
2N(N-1)\lambda &\leq Cp + C \sum_{k=2}^{N} (k-1) p(k).
\end{align}
Substituting $p(k) = {n \choose k} a_{N}^k \left( 1 - a_N\right)^{N-k}$ and computing the binomial sum we obtain
\begin{align}
2N(N-1)\lambda = N\left( 1 - \left(1 - a_N\right)^{N-1}\right).
\end{align}
Therefore,
\begin{align}
(N-1)\lambda &\leq \frac{1}{2} \left( 1 - \left(1 - a_N\right)^{N-1}\right), \\
&= \frac{1}{2}  \left( 1 - \left(1 - \frac{c}{N^\alpha}\right)^{N-1}\right).
\end{align}
When $0 < \alpha < 1$, we have $N/N^\alpha \rightarrow \infty$. In which case,
\begin{equation}
(N-1)\lambda \leq \frac{1}{2}  \left( 1 - \left(1 - \frac{c}{N^\alpha}\right)^{N-1}\right) = \Theta(1).
\end{equation}
Hence, $\lambda = O\left( 1/N\right)$. When $\alpha \geq 1$, either $N/N^\alpha \rightarrow 0$ or $N/N^\alpha \rightarrow c_1$ for some $c_1 > 0$. This implies
\begin{equation}
(N-1)\lambda \leq \frac{1}{2}  \left( 1 - \left(1 - \frac{c}{N^\alpha}\right)^{N-1}\right) = \Theta\left( N/N^\alpha\right).
\end{equation}
Hence, $\lambda = O\left( 1/N^\alpha\right)$.

\subsection{Proof of Expander Property and Flooding Time when $0 < \alpha < 1$}
\label{pf:expander_large}

\begin{framed}
\begin{lemma}
\label{lem:mid1}
For $1 \leq h \leq N^\alpha$
\begin{equation}
\label{eq:astro1}
\expect{N(h)} = \Theta\left( h N^{1 - \alpha} \right),
\end{equation}
and for all $N^{\alpha} + 1 \leq h \leq N/2$
\begin{equation}
\label{eq:astro2}
\expect{N(h)}  = \Theta(N).
\end{equation}
\end{lemma}
\end{framed}
\begin{IEEEproof}
We know that
\begin{equation}
N(h) \sim~\text{Bin}\left( N-h, 1 - \left( 1 - \frac{c}{N^\alpha}\right)^h \right). \nonumber
\end{equation}
Therefore,
\begin{equation}
\expect{N(h)} = (N-h)\left[ 1 - \left( 1 - \frac{c}{N^\alpha}\right)^h \right]. \nonumber
\end{equation}
If $h/N^\alpha \rightarrow 0$ then
\begin{equation}
\label{eq:tam1}
\frac{ 1 - \left( 1 - c/N^\alpha\right)^h }{ ch/N^\alpha } \rightarrow 1,
\end{equation}
and if $h/N^\alpha \rightarrow c_5$, for some $c_5 > 0$, then
\begin{equation}
\label{eq:tam2}
\frac{ 1 - \left( 1 - c/N^\alpha\right)^h }{ ch/N^\alpha } \rightarrow \frac{1 - \exp\{ - c c_5\}}{c c_5}.
\end{equation}
Since $f(x) = \frac{1 - \exp\{ - c x\}}{c x}$ is a decreasing function in $x$, from~\eqref{eq:tam1} and~\eqref{eq:tam2}
\begin{equation}
\frac{1 - e^{-c}}{c} \leq \lim_{N \rightarrow \infty} \frac{ 1 - \left( 1 - c/N^\alpha\right)^h }{ ch/N^\alpha } \leq 1, \nonumber
\end{equation}
for all $1 \leq h \leq N^\alpha$. This implies
\begin{equation}
\frac{1 - e^{-c}}{c} \leq \lim_{N \rightarrow \infty} \frac{ \expect{N(h)} }{ cN h/N^\alpha } \leq 1, \nonumber
\end{equation}
for all $1 \leq h \leq N^\alpha$. This proves~\eqref{eq:astro1}.

If $h/N^\alpha \rightarrow \infty$ then
\begin{equation}
\label{eq:tum1}
\lim_{N \rightarrow \infty} 1 - \left( 1 - c/N^\alpha\right)^h = 1,
\end{equation}
and if $h/N^\alpha \rightarrow c_6$, for some $c_6 > 0$, then
\begin{equation}
\label{eq:tum2}
\lim_{N \rightarrow \infty} 1 - \left( 1 - c/N^\alpha\right)^h = 1 - e^{-c c_6}.
\end{equation}
Since $f(x) = 1 - e^{-cx}$ is an increasing function of $x$, from~\eqref{eq:tum1} and~\eqref{eq:tum2} we have
\begin{equation}
1 - e^{-c} \leq \lim_{N \rightarrow \infty} 1 - \left( 1 - c/N^\alpha\right)^h \leq 1, \nonumber
\end{equation}
for all $N^{\alpha + 1} \leq h \leq N/2$. This implies
\begin{equation}
1 - e^{-c} \leq \lim_{N \rightarrow \infty} \frac{\expect{N(h)}}{N} \leq 1, \nonumber
\end{equation}
for all $N^{\alpha + 1} \leq h \leq N/2$. This proves~\eqref{eq:astro2}.
\end{IEEEproof}
Lemma~\ref{lem:mid1} implies that for all $h$, $\expect{N(h)} \rightarrow \infty$ as $N \rightarrow \infty$. Using Lemma~\ref{lem:binom_2} of Appendix~\ref{pf:concentration_bounds}, we have
\begin{equation}
\prob{N(h) < \eta_1 c_1 N^{1-\alpha} h } \leq_N \exp\left\{-c_5 N^{1-\alpha} h \right\}, \nonumber
\end{equation}
for some $\eta_1 \in (0,1)$, $c_5 > 0$, and for all $1 \leq h \leq N^\alpha$.\footnote{Note that the constant $c_5 > 0$ does not depend on $h$; see Lemma~\ref{lem:binom_2} in Appendix~\ref{pf:concentration_bounds}.} Using this and union bound, we obtain
\begin{align}
&\prob{\bigcup_{h=1}^{N^{\alpha}} \left\{ N(h) < \eta_1 c_1 N^{1 - \alpha} h \right\} } \nonumber \\
&~~~~~~~~~~~~~~\leq \sum_{h=1}^{N^\alpha} \prob{N(h) < \eta_1 c_1 N^{1 - \alpha} h }, \nonumber \\
&~~~~~~~~~~~~~~\leq \sum_{h=1}^{N^\alpha} \exp\left\{-c_5 N^{1-\alpha} h \right\}, \nonumber \\
&~~~~~~~~~~~~~~\leq \sum_{h=1}^{N^\alpha} \exp\left\{-c_5 N^{1-\alpha} \right\}, \nonumber \\
&~~~~~~~~~~~~~~= N^{\alpha} \exp\left\{-c_5 N^{1-\alpha} \right\} \leq_N \frac{c_6}{N^2},
\end{align}
for some $c_6 >0$.
This implies
\begin{equation}
\prob{\bigcap_{h=1}^{N^{\alpha}} \left\{ N(h) \geq \eta_1 c_1 N^{1-\alpha} h \right\} } \geq_N 1 - \frac{c_6}{N^2}.
\end{equation}
This proves the expander property for $1 \leq h \leq N^\alpha$.
Similarly, we obtain
\begin{equation}
\prob{\bigcap_{h=1}^{N^{\alpha}} \left\{ N(h) \geq \eta_2 c_3 N \right\} } \geq_N 1 - \frac{c_7}{N^2},
\end{equation}
for some $\eta_2 \in (0,1)$ and $c_7 > 0$, which is the expander property for $N^\alpha + 1 \leq h \leq N/2$. To prove~\eqref{eq:expander_large} we observe that if $\prob{A} \geq_N 1 - c_8/N^2$ and $\prob{B} \geq_N 1 - c_9/N^2$, for some positive constants $c_8$ and $c_9$, we have $\prob{A\cap B} \geq_N 1 - c_{10}/N^2$ for some positive constant $c_{10}$.

\subsubsection{Computing Flooding Time}
We now apply Theorem~\ref{thm:MEG_expander} to obtain an upper bound on flooding time.
\begin{multline}
\label{eq:zeza}
\sum_{h=1}^{N/2-1} \frac{ \log \left( \frac{h+1}{h}\right) }{ \log \left( 1 + k(h) \right) } = \sum_{h = 1}^{N^{\alpha}-1} \frac{\log \left( \frac{h+1}{h} \right)}{\log \left( 1 + c_1 N^{1-\alpha}\right)} \\
+ \sum_{h= N^\alpha}^{N/2-1} \frac{ \log \left( \frac{h+1}{h} \right) }{ \log \left( 1 + c_2 N/h \right)},
\end{multline}
for some $c_1 > 0$ and $c_2 > 0$. The first term in the expression can be simplified as
\begin{align}
\sum_{h = 1}^{N^{\alpha}-1} \frac{\log \left( \frac{h+1}{h} \right)}{\log \left( 1 + c_1 N^{1-\alpha}\right)} &=  \frac{\log \left( \prod_{h=1}^{N^\alpha-1} \frac{h+1}{h} \right)}{\log \left( 1 + c_1 N^{1-\alpha}\right)}, \\
&= \frac{\log N^\alpha}{ \log \left( 1 + c_1 N^{1-\alpha}\right) }, \\
&= \Theta\left( \frac{\log N^{\alpha}}{\log N^{1-\alpha}} \right) = \Theta(1). \label{eq:zeza2}
\end{align}
The second term in~\eqref{eq:zeza} can be simplified as
\begin{align}
\sum_{h= N^\alpha}^{N/2-1} \frac{ \log \left( \frac{h+1}{h} \right) }{ \log \left( 1 + c_2 N/h \right)} &\leq \sum_{h= N^\alpha}^{N/2-1} \frac{1}{h \log \left( 1 + c_2 N/h \right)}, \nonumber \\
&= \Theta\left( \int_{N^\alpha}^{N/2} \frac{dh}{h \log\left( 1 + c_2 N/h \right)} \right), \nonumber
\end{align}
where the first inequality is because $\log\left(1+\frac{1}{h}\right) \leq \frac{1}{h}$. To evaluate the integral, substitute $y = c_2 N/h$ to obtain
\begin{align}
&\int_{N^\alpha}^{N/2} \frac{dh}{h \log\left( 1 + c_2 N/h \right)} \nonumber \\
&~~~= \int_{2c_2}^{c_2 N^{1-\alpha}}\!\!\!\!\!\! \frac{dy}{y \log (1+y)},\nonumber \\
&~~~= \int_{2c_2}^{c_2 N^{1-\alpha}}\!\!\!\!\!\!\!\!\! \frac{1}{(1+y)}\frac{dy}{\log (1+y)} +  \int_{2c_2}^{c_2 N^{1-\alpha}}\!\!\!\!\!\!\!\!\! \frac{1}{y(1+y)}\frac{dy}{\log (1+y)}, \nonumber \\
&~~~\leq \left( 1 + \frac{1}{2c_2}\right)  \int_{2c_2}^{c_2 N^{1-\alpha}}\!\!\!\!\!\! \frac{1}{(1+y)}\frac{dy}{\log (1+y)}, \nonumber \\
&~~~= \left( 1 + \frac{1}{2c_2}\right) \left[\log \log\left( 1 + c_2 N^{1-\alpha}\right) - \log \log (2c_2)\right], \nonumber \\
&~~~= \Theta\left( \log \log N\right).
\end{align}
This implies
\begin{equation}
\sum_{h= N^\alpha}^{N/2-1} \frac{ \log \left( \frac{h+1}{h} \right) }{ \log \left( 1 + c_2 N/h \right)} = \Theta\left( \log \log N\right). \label{eq:zeza3}
\end{equation}
Hence, from~\eqref{eq:zeza},~\eqref{eq:zeza2}, and~\eqref{eq:zeza3}, the flooding time is upper bounded by $O(\log \log N)$.

\subsection{Proof of Expander Property and Flooding Time when $1 \leq \alpha < 2$}
\label{pf:expander_small}
Let $\beta >0$. We show that the network has expander property for $\beta N^{\alpha-1}\log N + 1 \leq h \leq N/2$ for some $\beta > 0$, and prove a lower bound on probability $\prob{N(h) \geq 1}$ for $1 \leq h \leq \beta N^{\alpha-1}\log N$.
\begin{framed}
\begin{lemma}
For every $\epsilon > 0$ we have
\begin{equation}
\prob{N(h) \geq 1} \geq_N 1 - (1+\epsilon)\exp\left\{-h/N^{\alpha - 1}\right\},
\end{equation}
for all $1 \leq h \leq \beta N^{\alpha -1} \log N$.
\end{lemma}
\end{framed}
\begin{IEEEproof}
Since $\prob{N(h) \geq 1} = 1 - \prob{N(h) = 0}$, we evaluate $\prob{N(h) = 0}$. We know that
\begin{equation}
N(h) \sim~\text{Bin}\left( N-h, 1 - \left(1 - c/N^\alpha\right)^h\right). \nonumber
\end{equation}
We thus have
\begin{equation}
\prob{N(h)=0} = \left( 1 - c/N^\alpha \right)^{h(N-h)}.
\end{equation}
This implies
\begin{equation}
\lim_{N \rightarrow \infty} \frac{\prob{N(h) = 0}}{  \exp\left\{ - c \frac{h(N-h)}{N^\alpha}\right\} } = 1.
\end{equation}
Note that $\frac{h(N-h)}{N^\alpha} = \frac{h}{N^{\alpha-1}} - \frac{h^2}{N^\alpha}$, and the first term in the expression dominates the scaling with $N$ for $1 \leq h \leq \beta N^{\alpha - 1}\log N$. Hence,
\begin{equation}
\lim_{N \rightarrow \infty} \frac{\prob{N(h) = 0}}{  \exp\left\{ - c Nh/N^\alpha \right\} } = 1,
\end{equation}
for all $1 \leq h \leq \beta N^{\alpha - 1}\log N$. This implies that for every $\epsilon > 0$
\begin{equation}
\prob{N(h) = 0} \leq_N (1+\epsilon)\exp\left\{-c h/N^{\alpha - 1}\right\},
\end{equation}
all $1 \leq h \leq \beta N^{\alpha-1}\log N$. This proves that for every $\epsilon > 0$
\begin{multline}
\prob{N(h) \geq 1} = 1 - \prob{N(h) = 0} \\
\geq_N 1 - (1+\epsilon)\exp\left\{-h/N^{\alpha - 1}\right\},
\end{multline}
all $h \in \{1, 2, \ldots \beta N^{\alpha -1} \log N\}$.
\end{IEEEproof}

\begin{framed}
\begin{lemma}
\label{lem:1}
For every $\epsilon > 0$ we have
\begin{equation}
(1-\epsilon)\frac{h}{N^{\alpha - 1}} \leq_N \expect{N(h)} \leq_N (1+\epsilon) \frac{h}{N^{\alpha - 1}},
\end{equation}
for all $\beta N^{\alpha - 1}\log N + 1 \leq h \leq N/2$.
\end{lemma}
\end{framed}
\begin{IEEEproof}
We know that
\begin{equation}
N(h) \sim~\text{Bin}\left( N-h, 1 - \left( 1 - \frac{c}{N^\alpha}\right)^h \right). \nonumber
\end{equation}
Therefore,
\begin{equation}
\expect{N(h)} = (N-h)\left[ 1 - \left( 1 - \frac{c}{N^\alpha}\right)^h \right].
\end{equation}
Note that if $h/N^\alpha \rightarrow 0$ then
\begin{equation}
\frac{ 1 - \left( 1 - c/N^\alpha\right)^h  }{ c h/N^\alpha } \rightarrow 1,
\end{equation}
and $h/N^\alpha \rightarrow 0$ for all $\beta N^{\alpha -1} \log N + 1 \leq h \leq N/2$. This implies
\begin{equation}
\lim_{N \rightarrow \infty} \frac{\expect{N(h)}}{c Nh/N^\alpha} = 1,
\end{equation}
for all $\beta N^{\alpha -1} \log N + 1 \leq h \leq N/2$. This proves the result.
\end{IEEEproof}

From Lemma~\ref{lem:1}, we note that $\expect{N(h)} \rightarrow \infty$ as $N \rightarrow \infty$ for all $\beta N^{\alpha - 1}\log N \leq h \leq N/2$. Using Lemma~\ref{lem:binom_2} of Appendix~\ref{pf:concentration_bounds}, we obtain for a given $\epsilon > 0$
\begin{equation}
\prob{N(h) < \eta (1-\epsilon)\frac{ch}{N^{\alpha - 1}} } \leq_N \exp\left\{ -c_1 \frac{ch}{N^{\alpha - 1}}\right\},
\end{equation}
for some $\eta \in (0,1)$, $c_1 > 0$, and all $h \in \{\beta N^{\alpha - 1}\log N + 1, \ldots \frac{N}{2} \}$.\footnote{Note that $c_1$ does not depend on $h$; see Lemma~\ref{lem:binom_2} in Appendix~\ref{pf:concentration_bounds}.} This, with union bound, implies
\begin{align}
&\prob{\bigcup_{h = \beta N^{\alpha - 1}\log N + 1}^{N/2} \left\{ N(h) < \eta(1-\epsilon)\frac{c h}{N^{\alpha - 1}} \right\} } \nonumber \\
&~~~\leq \sum_{h = \beta N^{\alpha - 1}\log N + 1}^{N/2} \prob{N(h) < \eta (1-\epsilon)\frac{ch}{N^{\alpha - 1}} }, \nonumber \\
&~~~\leq_N \sum_{h = \beta N^{\alpha - 1}\log N + 1}^{N/2} \exp\left\{ -c_1 \frac{ch}{N^{\alpha - 1}}\right\}, \nonumber \\
&~~~\leq N  \exp\left\{ -c_1 c \frac{ \beta N^{\alpha - 1}\log N + 1 }{N^{\alpha - 1}}\right\}, \nonumber \\
&~~~= \Theta\left( N \exp\left\{ -c_2 \beta \log N\right\} \right), \nonumber \\
&~~~= \Theta\left( \frac{1}{N^{c_2\beta - 1}} \right),
\end{align}
for some $c_2 > 0$. Choosing $\beta > 3/c_2$ we have
\begin{equation}
\prob{\!\!\bigcup_{h = \beta N^{\alpha - 1}\log N + 1}^{N/2}\!\!\! \left\{ N(h) < \eta (1-\epsilon)\frac{c h}{N^{\alpha - 1}} \right\} } \leq_N \frac{c_3}{N^2}, \nonumber
\end{equation}
for some $c_3 > 0$. This implies
\begin{equation}
\prob{\!\!\bigcap_{h = \beta N^{\alpha - 1}\log N + 1}^{N/2} \!\!\! \left\{ N(h) \geq \eta (1-\epsilon)\frac{c h}{N^{\alpha - 1}} \right\} } \geq_N 1  - \frac{c_3}{N^2}, \nonumber
\end{equation}
which proves the expander properties of~\eqref{eq:expander_small1}.

\subsubsection{Computing the Flooding Time}
Set
\begin{equation}
p(h) = 1 - c_4 \exp\left\{-ch/N^{\alpha - 1} \right\},
\end{equation}
for all $h \in \{1, 2, \ldots \beta N^{\alpha -1} \log N\}$ and some $c_4 > 0$. We know from Theorem~\ref{thm:MEG_hybrid} that the flooding time is upper bounded by
\begin{equation}
\label{eq:reff}
\sum_{h=1}^{\beta N^{\alpha -1} \log N} \frac{1}{p(h)} + \sum_{h=1}^{\beta N^{\alpha - 1}\log N - 1} \frac{ \log \left( \frac{h+1}{h}\right) }{ \log \left( 1 + c_5/N^{\alpha - 1}\right)},
\end{equation}
where $c_5 = \eta (1-\epsilon)c$.
Computing the first term we get
\begin{align}
\sum_{h=1}^{\beta N^{\alpha -1} \log N}\!\!\!\!\!\! \frac{1}{p(h)} &= \!\!\! \sum_{h=1}^{\beta N^{\alpha -1} \log N}\!\!\!\!\! \frac{1}{1 -  c_4 \exp\left\{-ch/N^{\alpha - 1} \right\} },\nonumber \\
&= \!\!\!\! \sum_{h=1}^{\beta N^{\alpha -1} \log N} \!\!\! \frac{ \exp\left\{ c h/N^{\alpha - 1} \right\} }{ \exp\left\{ c h/N^{\alpha - 1} \right\} - c_4 }, \nonumber \\
&= \Theta\left( \int_{1}^{ \beta N^{\alpha -1} \log N } \!\!\!\!\!\!\! \frac{ \exp\left\{ c h/N^{\alpha - 1} \right\} }{ \exp\left\{ c h/N^{\alpha - 1} \right\} - c_4 } dh \right). \nonumber
\end{align}
The integral equals
\begin{multline}
\int \frac{ \exp{ \left\{ c h/N^{\alpha -1} \right\} } }{ \exp{ \left\{ ch/N^{\alpha -1} \right\} } - c_4 } dh \\
= \frac{1}{c}N^{\alpha - 1} \log \left(  \exp{ \left\{ c h/N^{\alpha -1} \right\}} - c_4 \right). \nonumber
\end{multline}
We, thus, have
\begin{align}
\sum_{h=1}^{\beta N^{\alpha -1} \log N}\!\!\!\! \frac{1}{p(h)} &= \Theta\left( N^{\alpha - 1} \log \left(  \exp{ \left\{\beta \log N \right\} } - c_4 \right) \right), \nonumber \\
&= \Theta\left( N^{\alpha - 1} \log\left( N^{\beta} - c_4\right)\right), \nonumber \\
&= \Theta\left( N^{\alpha -1} \log N\right). \label{eq:temp1}
\end{align}
Computing the second term in the expression~\eqref{eq:reff} we have
\begin{align}
&\sum_{h=1}^{\beta N^{\alpha - 1}\log N - 1}\!\!\!\!\!\! \frac{ \log \left( \frac{h+1}{h}\right) }{ \log \left( 1 + c_5/N^{\alpha - 1}\right)} \nonumber \\
&~~~~~~~~~~~~~~~~= \frac{ \log \left( \prod_{h=1}^{\beta N^{\alpha - 1}\log N - 1} \frac{h+1}{h}\right) }{ \log \left( 1 + c_5/N^{\alpha - 1}\right)}, \nonumber \\
&~~~~~~~~~~~~~~~~= \frac{\log \left( \beta N^{\alpha - 1}\log N \right) }{ \log \left( 1 + c_5/N^{\alpha - 1}\right) }, \nonumber \\
&~~~~~~~~~~~~~~~~=\Theta\left( \frac{ \log N}{\log \left( 1 + c_5/N^{\alpha - 1}\right)}\right), \nonumber \\
&~~~~~~~~~~~~~~~~= \Theta\left( N^{\alpha - 1}\log N\right), \label{eq:temp2}
\end{align}
where the last equality follows because $\log\left(1+c_5/N^{\alpha - 1}\right) = \Theta\left( 1/ N^{\alpha - 1}\right)$. Therefore, from~\eqref{eq:temp1}, ~\eqref{eq:temp2}, and~\eqref{eq:reff} the flooding time is $T_N = O\left( N^{\alpha - 1} \log N\right)$ with probability at least $1 - c_6/N^2$ for some $c_6 > 0$.

\subsection{Proof of Expander Property and Flooding Time when $\alpha \geq 2$}
\label{pf:expander_tiny}
In this case, distribution of $N(h)$ is concentrated at $N(h) = 0$. We, therefore, seek a lower-bound on $\prob{N(h) = 1}$ in order to apply Theorem~\ref{thm:MEG_geobound}. Since
\begin{equation}
N(h) \sim \text{Bin}\left( N-h, 1 - \left( 1 - c/N^\alpha\right)^h\right),\nonumber
\end{equation}
we have
\begin{multline}
\prob{N(h) = 1} = (N-h)\left[ 1 - \left( 1 - \frac{c}{N^\alpha}\right)^h\right]\\
\times\left( 1 - \frac{c}{N^\alpha}\right)^{h(N-h-1)}, \label{eq:locala}.
\end{multline}
Note that $\frac{h}{N^\alpha} \rightarrow 0$ for all $h \in \{1, 2, \ldots, N-1\}$ since $\alpha \geq 2$. This implies
\begin{align}
\lim_{N \rightarrow \infty} \frac{1 - \left( 1 - \frac{c}{N^\alpha}\right)^h}{c h/N^\alpha} = 1, \label{eq:localb}
\end{align}
for all $h \in \{1, 2, \ldots N/2\}$. Also, since
\begin{equation}
\max_{h \in \{1, 2, \ldots N-1\} } h(N-h-1) \leq \frac{N^2}{4}, \nonumber
\end{equation}
and
\begin{equation}
\min_{h \in \{1, 2, \ldots N-1\} } h(N-h-1) \geq \frac{N}{2}, \nonumber
\end{equation}
we have
\begin{equation}
e^{-c/4} \leq \lim_{N \rightarrow \infty} \left( 1 - \frac{1}{N^\alpha}\right)^{h(N-h-1)} \leq 1. \label{eq:localc}
\end{equation}
Then, \eqref{eq:locala},~\eqref{eq:localb}, and~\eqref{eq:localc} imply
\begin{equation}
e^{-c/4} \leq \lim_{N \rightarrow \infty} \frac{\prob{N(h) = 1}}{(N-h)h/N^\alpha} \leq 1. \nonumber
\end{equation}
for all $1 \leq h \leq N-1$. Thus, there exists a positive constant $c_1$ such that
\begin{equation}
\prob{N(h) = 1} \geq_N c_1 \frac{(N-h)h}{N^\alpha}, \nonumber
\end{equation}
for all $1\leq h \leq N-1$. This proves the property of~\eqref{eq:expander_tiny} for
\begin{equation}
p(h) = c_1 \frac{(N-h)h}{N^\alpha}, \nonumber
\end{equation}
for all $1 \leq h \leq N-1$.

\subsubsection{Computing the Flooding Time}
Then the upper bound on flooding time given in Theorem~\ref{thm:MEG_geobound} equals
\begin{align}
\sum_{h=1}^{N-1}\frac{1}{p(h)} &=  \sum_{h=1}^{N-1} \frac{N^\alpha /c_1}{(N-h)h}, \nonumber  \\
&= \frac{1}{c_1}\frac{N^\alpha}{N} \sum_{h=1}^{N-1} \left[\frac{1}{h} + \frac{1}{N-h}\right], \nonumber \\
&= \Theta\left( N^{\alpha-1} \log N\right).
\end{align}

\subsection{Concentration Bounds}
\label{pf:concentration_bounds}
We list here some concentration bounds that we use in our proofs. The following Lemma is from Chap.~1 in~\cite{penrose_book}.
\begin{framed}
\begin{lemma}
\label{lem:binom_cbound}
If $X \sim \text{Bin}\left(n,p\right)$ for some $p \in (0,1)$ and $\mu = np$ then for all $k \geq \mu$
\begin{equation}
\prob{X \geq k} \leq \exp\left\{ -\mu H\left( \frac{k}{\mu}\right)\right\},
\end{equation}
and for all $k \leq \mu$
\begin{equation}
\prob{X \leq k} \leq \exp\left\{ -\mu H\left( \frac{k}{\mu}\right)\right\},
\end{equation}
where $H(a) = 1 - a + a\log a$ for all $a > 0$.
\end{lemma}
\end{framed}
We now extend this result to the following
\begin{framed}
\begin{lemma}
\label{lem:binom_2}
If $X_1, X_2, \ldots X_{g(n)}$ are binomial random variables such that
\begin{equation}
c_1 f(n) \leq_N \expect{X_h} \leq_N c_2 f(n),
\end{equation}
for some positive constants $c_1$ and $c_2$, where $g(n)$ and $f(n)$ are increasing functions of $n$. Then there exists an $\eta \in (0,1)$ and a positive constant $c_3$ such that
\begin{equation}
\prob{X_h < \eta c_1 f(n)} \leq_N e^{-c_3 f(n)},
\end{equation}
for all $h \in \{1, 2, \ldots g(n)\}$.
\end{lemma}
\end{framed}
%{\em Proof}:
\begin{IEEEproof}
For every $h \in \{1, 2, \ldots g(n)\}$, $X_h$ is a binomial random variable. Lemma~\ref{lem:binom_cbound} gives
\begin{equation}
\prob{X_h < \eta c_1 f(n)} \leq \exp \left\{ - \expect{X_h} H\left(\frac{\eta c_1 f(n)}{\expect{X_h}} \right)\right\}. \label{eq:abra}
\end{equation}
Evaluating the exponent of the right hand side, we get
\begin{align}
\expect{X_h}~&H\left(\frac{\eta c_1 f(n)}{\expect{X_h}}\right) \nonumber \\ &= \expect{X_h} - \eta c_1 f(n) + \eta c_1 f(n) \log\left( \frac{\eta c_1 f(n)}{\expect{X_h}}\right), \nonumber \\
&\geq_N c_1 f(n) - \eta c_1 f(n) + \eta c_1 f(n) \log\left( c_1 / c_2\right), \nonumber  \\
&=\left[ \frac{1-\eta}{\eta} - \log\left( c_2 / c_1\right)\right] \eta c_1 f(n).
\end{align}
where the second inequality follows from the fact that $c_1 f(n) \leq_n \expect{X_h} \leq_n c_2 f(n)$. Now, since $\frac{1-\eta}{\eta}$ can take any positive real values for $\eta \in (0,1)$, we have
\begin{equation}
\expect{X_h} H\left(\frac{\eta c_1 f(n)}{\expect{X_h}}\right) \geq c_3 f(n), \label{eq:dabra}
\end{equation}
for some $\eta \in (0,1)$ and $c_3 = \left[ \frac{1-\eta}{\eta} - \log\left( c_2 / c_1\right)\right] \eta c_1 > 0$ for the corresponding $\eta$. Notice that $c_3$ does not depend on $h$, and hence,~\eqref{eq:dabra} holds for all $h \in \{1, 2, \ldots g(n)\}$. Combining~\eqref{eq:abra} and~\eqref{eq:dabra} we obtain
\begin{equation}
\prob{X_h < \eta c_1 f(n)} \leq_n \exp\left\{ -c_3 f(n)\right\},
\end{equation}
for all $h \in \{1, 2, \ldots g(n)\}$.%~\qed
\end{IEEEproof}

\begin{framed}
\begin{lemma}
\label{lem:geo_cbound}
Let $X_1, X_2, \ldots X_n$ be independent geometrically distributed random variables with parameters $0 < p_1 \leq p_2 \leq \cdots \leq p_n$, i.e., $\prob{X_i = t} = p_i(1-p_i)^{t-1}$ for all $t \geq 1$. Let $S_n = \sum_{i=1}^{n} X_i$ and
\begin{equation}
\mu = \expect{S_n} = \frac{1}{p_1} + \frac{1}{p_2} + \cdots + \frac{1}{p_n}.
\end{equation}
Then, for some $c \geq 2$,
\begin{equation}
\prob{S_n > c(\mu + t) } \leq (1-p_1)^t \exp\left\{-(2c-3)n/4 \right\}.
\end{equation}
\end{lemma}
\end{framed}
\begin{IEEEproof}
The proof is given in~\cite{tailbound}.
\end{IEEEproof}

%%%%%%%%%%%%%%%%%%%%%%%%%%%%%%%%%%%%%%%%%%%%%%%%%%
%\bibliographystyle{IEEEtran}
\bibliographystyle{ieeetr}

\end{document}